\documentclass[fleqn,twoside]{article}
\usepackage{espcrc2}

\usepackage{graphicx}
\usepackage[figuresright]{rotating}

\newcommand{\be}{\begin{equation}}
\newcommand{\ee}{\end{equation}}
\newcommand{\ba}{\begin{eqnarray}}
\newcommand{\ea}{\end{eqnarray}}
\newcommand{\dis}{\displaystyle}

\newcommand{\AmS}{{\protect\the\textfont2
  A\kern-.1667em\lower.5ex\hbox{M}\kern-.125emS}}

\title{$|V_{us}|$ and $m_s$ from hadronic tau decays}

\author{Elvira~G\'amiz\address{Department of Physics, University of Illinois,
Urbana IL 61801, USA},
Matthias~Jamin\address{ICREA and IFAE, Departament de F\'{\i}sica Te\`orica,
Universitat Auton\`oma de Barcelona,\\ E-08193 Bellaterra, Barcelona, Spain},
\thanks{Email: jamin@ifae.es; Invited talk given by M.J.
at ``Tau06, 19-22 September 2006, Pisa, Italy''.}
Antonio~Pich\address{Departament de F\'{\i}sica Te\`orica, IFIC,
Universitat de Val\`encia--CSIC \\  Apt. de Correus 22085, E-46071
Val\`encia, Spain},
Joaquim~Prades\address{Theory Unit, Physics Department, CERN, CH-1211 Gen\`eve 23,
Switzerland},\thanks{On leave of absence from CAFPE and Departamento de
F\'isica Te\'orica y del Cosmos, Universidad de Granada, Campus de Fuente
Nueva, E-18002 Granada, Spain.}
Felix~Schwab\address{IFAE, Departament de F\'{\i}sica Te\`orica,
Universitat Auton\`oma de Barcelona,\\ E-08193 Bellaterra, Barcelona, Spain}}

\begin{document}

\begin{abstract}
\vspace{-12cm}
\begin{flushright}
CERN-PH-TH/2006-254, UAB-FT-620\\
\end{flushright}
\vspace{11cm}

Recent progress in the determination of $|V_{us}|$ employing strange hadronic
$\tau$-decay data are reported. This includes using the recent OPAL update of
the strange spectral function, as well as augmenting the dimension-two
perturbative contribution with the recently calculated order $\alpha_s^3$
term on the theory side. These updates result in $|V_{us}|=0.2220 \pm 0.0033$,
with the uncertainty presently being dominated by experiment, and already being
competitive with the standard extraction from $K_{e3}$ decays and other new
proposals to determine $|V_{us}|$. In view of the ongoing work to analyse
$\tau$-decay data at the B-factories BaBar and Belle, as well as future
results from BESIII, the error on $|V_{us}|$ from $\tau$ decays is expected to
be much reduced in the near future.
\end{abstract}

\maketitle

\section{INTRODUCTION}

In the past decade hadronic $\tau$ decays have been an extremely fruitful
laboratory for the study of low-energy QCD. Detailed investigations of the
$\tau$ hadronic width
\begin{equation}
R_\tau \,\equiv\, \frac{\Gamma[\tau^- \to {\rm hadrons} \, \nu_\tau (\gamma)]}
{\Gamma[\tau^- \to e^- \overline \nu_e \nu_\tau (\gamma)]} \,,
\end{equation}
as well as invariant mass distributions, have served to determine the QCD
coupling $\alpha_s$ to a precision competitive with the current world average
\cite{BNP92,CLEO95,ALEPH98,OPAL99,ALEPH05}.
The experimental separation of the Cabibbo-allowed decays and
Cabibbo-suppressed modes into strange particles
\cite{ALEPH99,CLEO03,OPAL04,CHD97,DHZ06} also opened a means to determine the
mass of the strange quark \cite{PP98,CKP98,PP99,KKP00,KM00,CDGHPP01,GJPPS03},
one of the fundamental QCD parameters within the Standard Model.

These determinations suffer from large QCD corrections to the contributions of
scalar and pseudoscalar correlation functions \cite{BNP92,PP98,CK93,Mal98a}
which are additionally amplified by the particular weight functions which
appear in the $\tau$ sum rule. A natural remedy to circumvent this problem is
to replace the QCD expressions of scalar and pseudoscalar correlators by
corresponding phenomenological hadronic parametrisations
\cite{ALEPH99,PP99,KM00,MK01,GJPPS03}, which turn out to be more precise
than their QCD counterparts, since the dominant contribution stems from
the well known kaon pole.

Additional suppressed contributions to the pseudoscalar correlators come from
the pion pole as well as higher exited pseudoscalar states whose parameters
have recently been estimated \cite{MK02}. The remaining strangeness-changing
scalar spectral function has been extracted from a study of S-wave $K\pi$
scattering \cite{JOP00,JOP01} in the framework of resonance chiral
perturbation theory \cite{EGPR89}. The resulting scalar spectral function was
also employed to directly determine $m_s$ from a purely scalar QCD sum rule
\cite{JOP02}.

Nevertheless, as was already realised in the first works on strange mass
determinations from the Cabibbo-suppressed $\tau$ decays, $m_s$ turns out
to depend sensitively on the element $|V_{us}|$ of the quark-mixing (CKM)
matrix. With the theoretical improvements in the $\tau$ sum rule mentioned
above, in fact $|V_{us}|$ represents one of the dominant uncertainties for
$m_s$. Thus it appears natural to actually determine $|V_{us}|$ with an input
for $m_s$ as obtained from other sources \cite{GJPPS03,JAM03,GJPPS05}.

Succeeding the high-precision status on $\tau$-decay observables already
attained by by ALEPH and OPAL at LEP and CLEO at CESR, now the B-factories
BaBar and Belle are starting to produce their first results on hadronic
$\tau$ decays, and in particular on Cabibbo-suppressed modes
\cite{NUG06,SCH06,OHS06}. These two facts make the strange hadronic $\tau$
decay data an ideal place for determining SU(3) breaking parameters such as
$|V_{us}|$ and/or $m_s$. The obvious advantage of this procedure is that the
experimental uncertainty will eventually be reduced at the B-factories and at
future facilities like the $\tau$-charm factory BEPCII 

\section{THEORETICAL FRAMEWORK}

Employing the analytic properties of two-point correlation functions for
vector ($\mathcal{J}=V$) and axial-vector ($\mathcal{J}=A$) two
quark-currents,
\begin{eqnarray}
\Pi^{\mu\nu}_{\mathcal{J}, ij}(q) &\!\!\!\equiv\!\!\!& i \int {\rm d}^4 x \,
e^{i q \cdot x} \langle 0 | T [ \mathcal{J}_{ij}^{\mu\dagger}(x)
\mathcal{J}^\nu_{ij}(0) ]| 0 \rangle \nonumber \\
&& \hspace{-15mm} \equiv \left[q^\mu q^\nu - q^2 g^{\mu\nu}\right]
\Pi^T_{\mathcal{J}, ij}(q^2) + q^\mu q^\nu \Pi^L_{\mathcal{J},
ij}(q^2) \,, \nonumber \\ &&
\end{eqnarray}
one can express $R_\tau$  as a contour integral running counter-clockwise
around the circle $|s|=M_\tau^2$ in the complex s-plane:
\begin{eqnarray}
R_\tau &\!\!\!\equiv\!\!\!& -\,i \pi \oint_{|s|=M_\tau^2}
\frac{{\rm d} s}{s} \, \left[1-\frac{s}{M_\tau^2}\right]^3 \\
&\!\!\!\times\!\!\!& \left\{ 3 \left[1+\frac{s}{M_\tau^2}
\right] D^{L+T}(s) + 4 \, D^L(s) \right\} \,. \nonumber
\end{eqnarray}
Here, we have used integration by parts to rewrite $R_\tau$ in terms of
the logarithmic derivatives
\begin{eqnarray}
D^{L+T}(s) &\!\!\!\equiv\!\!\!& -\,s \, \frac{{\rm d}}{{\rm d } s }
\, \Pi^{L+T}(s) \,, \nonumber \\
D^{L}(s) &\!\!\!\equiv\!\!\!& \frac{s}{M_\tau^2} \, \frac{{\rm d} }{{\rm d } s}
\, \left[ s \Pi^{L}(s) \right] \, \,.
\end{eqnarray}
Moreover, experimentally $R_\tau$ can be decomposed into the following three
contributions
\begin{equation}
R_\tau \,=\, R_{\tau, V} + R_{\tau, A} + R_{\tau, S} \,,
\end{equation}
according to the quark content
\begin{eqnarray}
\Pi^J(s) &\!\!\!=\!\!\!& |V_{ud}|^2 \left\{ \Pi^J_{V,ud}(s) +
\Pi^J_{A,ud}(s) \right\} \nonumber \\
&\!\!\!+\!\!\!& |V_{us}|^2 \left\{ \Pi^J_{V,us}(s) + \Pi^J_{A,us}(s) \right\}
\,,
\end{eqnarray}
where $R_{\tau, V}$ and $R_{\tau, A}$ correspond to the first two terms in
the first line and $R_{\tau, S}$  to the second line, respectively.

Additional information can be inferred from the measured invariant-mass
distribution of the final hadrons, which defines the moments
\begin{equation}
\label{OPE} R_\tau^{kl} \equiv {\dis \int^{M_\tau^2}_0}
{\rm d} s \left(1-\frac{s}{M_\tau^2}\right)^k \left( \frac{s}{M_\tau^2}
\right)^l \, \frac{{\rm d} R_\tau}{{\rm d} s} \,.
\end{equation}
At large enough Euclidean $Q^2\equiv -s$, both $\Pi^{L+T}(Q^2)$ and
$\Pi^{L}(Q^2)$ can be organised in a dimensional operator series using well
established QCD operator product expansion (OPE) techniques. One then obtains
\begin{eqnarray}
R_\tau^{kl} &\!\!\!=\!\!\!& N_c \, S_{\rm EW} \Big\{ (|V_{ud}|^2 +
|V_{us}|^2) \,  \left[ 1 + \delta^{kl(0)}\right] \nonumber \\
 && +\,{\dis \sum_{D\geq2}} \left[ |V_{ud}|^2 \delta^{kl(D)}_{ud}
+ |V_{us}|^2 \delta^{kl(D)}_{us} \right] \Big\} \,.
\end{eqnarray}
The electroweak radiative correction $S_{\rm EW}=1.0201\pm0.0003$ \cite{Sew}
has been pulled out explicitly and $\delta^{kl(0)}$ denote the purely
perturbative dimension-zero contributions.  The symbols $\delta^{kl(D)}_{ij}$
stand for higher dimensional corrections in the OPE from dimension $D\geq 2$
operators, which contain  implicit $1/M_\tau^D$ suppression factors
\cite{BNP92,PP98,PP99,CK93}. The most important being the operators $m_s^2$
with $D=2$  and $m_s \langle \overline q q \rangle$ with $D=4$.

 In addition, the flavour-SU(3) breaking quantity
\begin{eqnarray}
\delta R^{kl}_\tau &\!\!\!\equiv\!\!\!& \frac{R^{kl}_{\tau, V+A}}{|V_{ud}|^2} -
\frac{R^{kl}_{\tau, S}}{|V_{us}|^2} \nonumber \\
&\!\!\!=\!\!\!& N_c \, S_{EW} \, {\dis \sum_{D\geq 2}} \left[\,
\delta^{kl(D)}_{ud}-\delta^{kl(D)}_{us}\right]
\end{eqnarray}
enhances the sensitivity to the strange quark mass. The dimension-two
correction $\delta^{kl(2)}_{ij}$ is known to order $\alpha_s^3$ for both
correlators, $J=L$ and $J=L+T$ \cite{PP98,PP99,BCK05}.

In Ref.~\cite{PP98}, an extensive analysis of this $D=2$ correction was
performed and it was shown that the perturbative $J=L$ correlator behaves
very badly. The $J=L+T$ correlator was also analysed there to order
$\alpha_s^2$ and showed a relatively good convergence. In the following, we
have included the recently calculated $O(\alpha_s^3)$ correction for $J=L+T$
\cite{BCK05}. One can see that the $J=L+T$ series starts to show its asymptotic
character at this order, though it is still much better behaved than the $J=L$
component. Due to the asymptotic behaviour, it does not make much sense to
sum all known orders of the series. This question will be investigated in more
detail by us in the future.

\section{DETERMINATION OF \boldmath{$|V_{us}|$} WITH FIXED \boldmath{$m_s$}}

One can now use the  relation
\begin{equation}
\label{Vus}
|V_{us}|^2 = \frac{R_{\tau, S}^{00}}{\dis \frac{R_{\tau,
V+A}^{00}}{|V_{ud}|^2}-\delta R_{\tau, {\rm th}}^{00}} \,,
\end{equation}
and analogous relations for other moments, to determine the
Cabibbo-Kobayashi-Maskawa (CKM) matrix element $|V_{us}|$. Notice that
on the right-hand side of (\ref{Vus}) the only theoretical input is
$\delta R_{\tau, {\rm th}}^{00}$, which is around 0.24 and should be
compared to the experimental quantity $R_{\tau, V+A}^{00}/|V_{ud}|^2$ which
is around 3.7. Therefore, with a not so precise theoretical prediction for
$\delta R_{\tau, {\rm th}}^{00}$ one can get a quite accurate value for
$|V_{us}|$, depending on the uncertainty provided by experiment.

The very bad QCD behaviour of the $J=L$ component in
$\delta R_{\tau, {\rm th}}^{kl}$ induces a large theoretical uncertainty,
which can be reduced considerably using phenomenology for the scalar and
pseudo-scalar correlators \cite{GJPPS03,JAM03,GJPPS05}. In particular, the
pseudo-scalar spectral functions are dominated by far by the well-known kaon
pole, to which we add suppressed contributions from the pion pole, as well
as higher excited pseudo-scalar states whose parameters have been estimated
in Ref.~\cite{MK02}. For the strange scalar spectral function, we employ the
result \cite{JOP00}, obtained from a study of S-wave $K\pi$ scattering within
resonance chiral perturbation theory \cite{EGPR89}, which has been recently
updated in Ref.~\cite{JOP06}.

The smallest theoretical uncertainty arises for the $kl=00$ moment, for
which we get
\begin{eqnarray}
\label{deltaR}
\delta R_{\tau, {\rm th}}^{00} &\!\!\!=\!\!\!& 0.1544\,(37) +
9.3\,(3.4) \, m_s^2 \nonumber \\
&\!\!\!+\!\!\!& 0.0034\,(28)\,=\, 0.240\,(32) \,,
\end{eqnarray}
where $m_s$ denotes the strange quark mass in units of GeV, and in the
$\overline{\mathrm{MS}}$ scheme at a renormalisation scale of $\mu=2$ GeV.
The first term contains the phenomenological scalar and pseudo-scalar
contributions, the second term contains the rest of the perturbative $D=2$
contribution, while the last term stands for the rest of the contributions.
Notice that the phenomenological contribution is more than 64\% of the total,
while the rest comes almost from the perturbative $D=2$ contribution. Here,
we update $\delta R_{\tau, {\rm th}}^{00}$ of Refs.~\cite{GJPPS03,JAM03,GJPPS05}
in various respects. Firstly, we use the recently updated scalar spectral
function \cite{JOP06}; secondly, we include the $\alpha_s^3$ corrections to
the $J=L+T$ correlator as calculated in Ref.~\cite{BCK05}; and finally, we use
an average of contour improved \cite{DP92} and fixed order perturbation results
for the asymptotically summed series, in order to have a more conservative
account of uncertainties resulting from unknown higher orders. A detailed
discussion of this contribution will be presented elsewhere \cite{GJPPS}.

For the $m_s$ input value, we use the recent average
$m_s(2\,{\rm GeV}) = (94 \pm 6)$ MeV \cite{JOP06}, which includes the most
recent determinations of $m_s$ from QCD sum rules and lattice QCD. The
strange quark mass uncertainty corresponds to the most precise determination
from the lattice.

Recently, Maltman and Wolfe have criticised the theory error we previously
employed for the $D=2$ OPE coefficient \cite{MW06}. Awaiting a more detailed
study \cite{GJPPS}, in our updated estimate (\ref{deltaR}), we have decided
to include a more conservative estimate of unknown higher-order corrections
by using an average of contour improved and fixed-order perturbation theory.
Still, we do not think that artificially doubling the perturbative uncertainty,
as was done in \cite{MW06}, represents an error estimate which is better
founded. Notice furthermore, that $\delta R_{\tau, {\rm th}}^{00}$ is dominated
by the scalar and pseudoscalar contributions which are rather well known from
phenomenology, and that the larger perturbative uncertainty is compensated by
the smaller $m_s$ error, so that our final theoretical uncertainty is
practically the same as in previous works \cite{GJPPS03,JAM03,GJPPS05}.

In order to finally determine $|V_{us}|$, we employ the following updates of
the remaining input parameters: $|V_{ud}|=0.97377 \pm 0.00027$ \cite{PDG06},
the non-strange branching fraction $R_{\tau, V+A}^{00}= 3.471 \pm 0.011$,
,\cite{DHZ06} as well as the strange branching fraction
$R_{\tau, S}^{00} = 0.1686 \pm 0.0047$ \cite{DHZ06} (see also
Refs.~\cite{ALEPH99} and \cite{OPAL04}), which includes the theoretical
prediction for the decay $B[\tau \to K \nu_\tau (\gamma)] = 0.715 \pm 0.003$
which is based on the better known $K\to \mu \nu_\mu(\gamma)$ decay rate.
For $|V_{us}|$, we then obtain
\begin{equation}
\label{numVus}
|V_{us}| = 0.2220 \pm 0.0031_{\rm exp} \pm 0.0011_{\rm th} \,.
\end{equation}
The experimental uncertainty includes a small component from the error in
$|V_{ud}|$, but it is dominated by the uncertainty in $R_{\tau, S}^{00}$,
while the theoretical error is dominated by the uncertainty in the
perturbative expansion of the $D=2$ contribution.

\section{SIMULTANEOUS FIT OF \boldmath{$|V_{us}|$} AND \boldmath{$m_s$}}

In principle, it is also possible to perform a simultaneous fit to $|V_{us}|$
and $m_s$ from a certain set of $(k,l)$ moments. As soon as more precise data
are available, this will be the ultimate approach to determine $|V_{us}|$ and
$m_s$ from hadronic $\tau$ decays. With the current uncertainties in the data
and a persistent question about a monotonous $k$-dependence of $m_s$
\cite{GJPPS03,GJPPS05}, a bias could be present in the method. Furthermore,
the correlations between different moments are rather strong and also have to
be properly included on the theory side.

Here, we shall restrict ourselves to a simplified approach where all
correlations are neglected. For the simultaneous fit of $|V_{us}|$ and $m_s$,
we employ the five $R_\tau^{kl}$ moments $(0,0)$ to $(4,0)$ which have also
been used in our previous analyses \cite{GJPPS03,GJPPS05}. Performing this
exercise, for the central values we find:
\begin{equation}
\label{msfit}
|V_{us}| = 0.2196 \,, \qquad m_s(2\,{\rm GeV}) = 76 \, {\rm MeV} \,.
\end{equation}
The expected uncertainties on these results should be smaller than the
individual error given in eq.~(\ref{numVus}) and the one for $m_s$ presented
in Ref.~\cite{GJPPS05}, but only slightly since the correlations between
different moments are rather strong.

The general trend of the fit result can be understood easily. $m_s$ from
the simultaneous fit turned out lower than the global average
$m_s(2\,{\rm GeV})=94\pm6\,{\rm MeV}$ considered above. Thus, also the
corresponding $\delta R_{\tau,{\rm th}}$ is lower, resulting in a reduction
of $|V_{us}|$. Furthermore, the moment-dependence of $m_s$ is reduced as
compared to our analysis \cite{GJPPS03} on the basis of the ALEPH data alone.
Nevertheless, we shall leave a detailed error analysis for a future publication.

\section{CONCLUSIONS}

High precision Cabibbo-suppressed hadronic $\tau$ data from ALEPH and OPAL at
LEP and CLEO at CESR already provide a competitive result for  $|V_{us}|$. As
presented above and in Refs.~\cite{GJPPS03,JAM03,GJPPS05}, the final
uncertainty in the $\tau$ determination of $|V_{us}|$ becomes an experimental
issue and will eventually be much reduced with the new B-factory data
\cite{NUG06,SCH06,OHS06}, and further reduced at future $\tau$ facilities.
A combined fit to determine both $|V_{us}|$ and $m_s$ will then be possible.
Hadronic $\tau$ decays have the potential to provide the most accurate
measurement of $|V_{us}|$ and a very competitive $m_s$ determination.

\section*{Acknowledgements}
This work has been supported in part by the European Commission (EC) RTN
FLAVIAnet under Contract No. MRTN-CT-2006-035482, 
by MEC (Spain) and FEDER (EC) Grant Nos. FPA2005-02211 (M.J.),
FPA2004-00996 (A.P.) and FPA2006-05294 (J.P.), by Junta de Andaluc\'{\i}a
Grant Nos. P05-FQM-101 (J.P.) and P05-FQM-437 (E.G. and J.P.) and by the
Deutsche Forschungsgemeinschaft (F.S.).


\end{document}